# EMBEDDINGS OF LINEAR ARRAYS, RINGS AND 2-D MESHES ON EXTENDED LUCAS CUBE NETWORKS

## Ernastuti[1], Vincent Vajnovzki[2]

[1]Department of Computer Science, Gunadarma University
Jalan Margonda Raya 100, Depok, Indonesia
E-mail : ernas@staff.gunadarma.ac.id

[2]Universite de Bourgogne, France
E-mail : vvajnov@u-bourgogne.fr

**Abstract**

A Fibonacci string is a length *n* binary string containing no two consecutive 1s. Fibonacci cubes (FC), Extended Fibonacci cubes (ELC) and Lucas cubes (LC) are subgraphs of *hypercube* defined in terms of Fibonacci strings. All these cubes were introduced in the last ten years as models for interconnection networks and shown that their network topology posseses many interesting properties that are important in parallel processor network design and parallel applications. In this paper, we propose a new family of Fibonacci-like cube, namely *Extended Lucas Cube* (ELC). We address the following network simulation problem : Given a linear array, a ring or a two-dimensional mesh; how can its nodes be assigned to ELC nodes so as to keep their adjacent nodes near each other in ELC ?. We first show a simple fact that there is a Hamiltonian path and cycle in any ELC. We prove that any linear array and ring network can be embedded into its corresponding optimum ELC (the smallest ELC with at least the number of nodes in the ring) with dilation 1, which is optimum for most cases. Then, we describe dilation 1 embeddings of a class of meshes into their corresponding optimum ELC.

***Keywords :*** (Extended) Fibonacci cube, Extended Lucas cube, Fibonacci number, Hamiltonian path, Hamiltonian cycle, linear array, ring , mesh, network.

## 1. Introduction

A widely studied network topology is the *Hypercube* [1]. The hypercube provides a rich network structure which permits many other topologies to be efficiently emulated. Most other popular networks are easily embedded into a hypercube. For example a 8 node linear array, 2x4 mesh and 8 node ring may be embedded into a 8 node hypecube as shown in Figure 1. It is well known that any hypercube is Hamiltonian. Clearly, there must be an embedding of a linear array and ring into Hypercube. Furthermore, the number of nodes $2^n$ in an n-dimensional hypercube H(n) grows rapidly as n increases [2]. This limits considerably the choice of the number of nodes in the graph. Figure 1 shows Hypercube of dimension n, n = 0,1,...,4.

The *Fibonacci cube* (FC) proposed by Hsu [3] is a special subcube of a hypercube based on Fibonacci numbers.

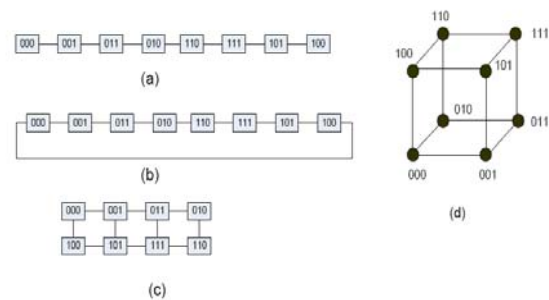

**Fig 1. a) Linear Array-8 node, b) Ring-8 node, c) 2-D Mesh(2x4), d) Hypercube-8 node**

Because of the rich peroperties of Fibonacci numbers, this network shows interesting properties. It has been shown that the diameter and node degree of the Fibonacci cube network with N nodes are O( log N), which are similar to those of the Hypercube of O(N) nodes. The Fibonacci cube has a very attractive recurrent structure, as it can be decomposed into networks which are also Fibonacci cubes. It has also been shown that the Fibonacci cube contains about 1/5 fewer edges than the hypercube for the same number of nodes. The size of the Fibonacci cube does not increase as fast as the hypercube with many faulty nodes.

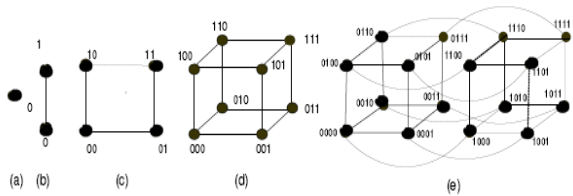

**Fig.2 . a) H(0), b) H(1), c)H(2), d) H(3), e) H(4)**

Throughout the paper, the term "graph" and term "network" means the same and are used interchangeably. Most Fibonacci cube are not Hamiltonian. In fact, Cong et al, showed that less than 1/3 of Fibonacci cube are Hamiltonian. Furthermore, Wu [1] proposed the *Extended Fibonacci Cube* (EFC), which is based on the same sequence as the Fibonacci cube; however its initial conditions are defined diffrently. It has been shown that EFC is Hamiltonian and maintains virtually all the desirable properties of FC. It has been shown that EFC is superior to FC in terms of various structural properties. All EFC are Hamiltonian. While Munarini et al proposed an other topology called *Lucas Cube* (LC), which is an induced subgraph of FC. LC is more sparse than both FC and EFC . LC are not Hamiltonian, however, almost all of the LC have a Hamiltonian path. The existance of a Hamiltonian path or cycle in cubes is a proof that linear array and ring can be embedded into the cubes. Network embedding has been studied extensively [4]. Thus far, there is a big gap between other networks (such as hypercube, star graph, ets) and the FC in relation to the availability of routing and network embedding results. The problem of embedding meshes into the Fibobacci cube was studied. It was shown that the Fibonacci cube contains two small meshes. This result did not take into consideration the problem of embedding meshes into their optimum Fibonacci cube, which is a more interesting problem. However, furthermore Cong showed that the meshes can be embedded into their optimum Fibonacci cube.

In this paper, we propose a new family of Fibonacci-like cube, namely *Extended Lucas Cube* (ELC). ELC is a network model obtained from EFC by removing nodes which are not Lucas string. Thus ELC is induced subgraph of EFC . In [5] can be found about Lucas string.

An embedding is defined as a mapping function g, which maps the vertices of a guest graph G to the vertices of a host graph H. The dilation of an embedding g is the maximum distance in H between images of the adjacent nodes in G. The expansion of an embedding g is the ratio of the number of nodes in H to the number of the nodes in G. The dilation is a lower bound for communication delay when H simulates G. The expansions are measurements for processor utilization of H. The problem of interest is : how to map the nodes of a ring and a two dimensional mesh to the nodes of its corresponding optimum ELC (the smallest ELC with at least as many nodes as the ring or mesh), on a one-to-on basis, so that dilation is kept minimum.

## 2. Network Topology

A Fibonacci cube of dimension n, denoted by FC(n), is an undirected graph of $f_n$ (which is the n-th Fibonacci number) nodes, each is labeled by a distinct (n–2)-bit binary number such that no two 1's occur consecutively. Two nodes in FC(n) are connected by an edge if and if their labels differ in exactly one bit position. Figure 3 shows FC(3), FC(4), FC(5) and FC(6). We use $|FC(n)|$ to denote the number of nodes in FC(n). The following are some basic properties of FC(n):

1. Assume FC(n) = (V(n),E(n)). FC(n-1)=(V(n-1),E(n-1)), and FC(n-2)=(V(n-2),E(n-2)), then V(n) = 0V(n-1)∪10V(n-2). The initial conditions are V(2)={} and V(3)={0,1}. For example, V(5) = 0{00,01,10} ∪ 10{0,1} = {000,001,010,100,101}. In other words, FC(n) can be decomposed into two subgraphs that are isomorphic to FC(n-1) and FC(n-2), respectively, for n ≥ 4. Figure 3 illustrates FC(n), n=3,4,5,6.
2. Let H(n) denote the hypercube of dimension n. Then FC(n) is a subgraph of H(n-2) and H(n) is a subgraph of FC(2n+1).

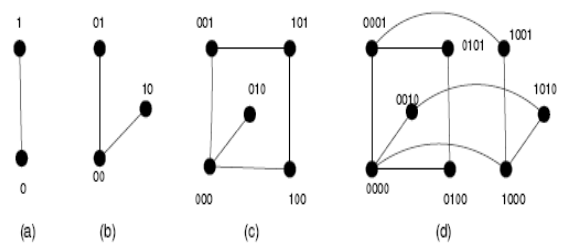

**Fig. 3. a) FC(3), b) FC(4), c) FC(5) , d) FC(6)**

An extended Fibonacci cube of dimension n, denoted by EFC(n), is an undirected graph of $F_n$ (which is $2.f_{n-1}$) nodes [6], each is labeled by a distinct (n–2)-bit binary number. Two nodes in FC(n) are connected by an edge if and if their labels differ in exactly one bit position. Figure 4 shows EFC(3), EFC(4), EFC(5) and EFC(6). We use $|EFC(n)|$ to denote the number of nodes in EFC(n). The following are some basic properties of EFC(n):

1. The two initial values are 2 and 4.
2. Assume EFC(n) = $(V_1(n),E_1(n))$. EFC(n-1)=$(V_1(n-1),E_1(n-1))$, and EFC(n-2)=$(V_1(n-2),E_1(n-2))$, then $V_1(n)= 0V_1(n-1) \cup 10V_1(n-2)$. The initial conditions are V(3)={0,1} and V(4)={00,10,11,01}. For example, V(5) = 0{00,10,11,01} $\cup$ 10{0,1} = {000,010,011,001,100,101}. In other words, EFC(n) can be decomposed into two subgraphs that are isomorphic to EFC(n-1) and EFC(n-2), respectively, for n $\geq$ 5. Figure 4 illustrates EFC(n), n=3,4,5,6.

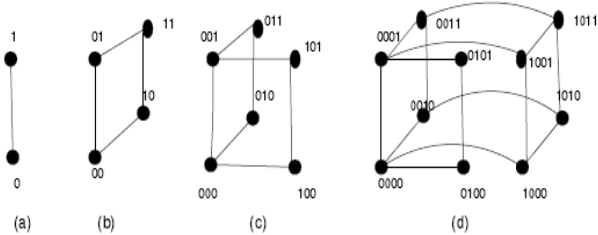

**Fig. 4. a) EFC(3), b) EFC(4), c) EFC(5) , d) EFC(6)**

An extended Lucas cube of dimension n, denoted by ELC(n), is an undirected graph of $C_n$ (which is $2.f_n \times 2.f_{n-2}$) nodes [7] , each is labeled by a distinct (n–2)-bit binary number. Two nodes in ELC(n) are connected by an edge if and if their labels differ in exactly one bit position. Figure 5 shows ELC(3), ELC(4), ELC(5) and ELC(6). We use $|ELC(n)|$ to denote the number of nodes in ELC(n).

Assume ELC(n) = $(V_2(n),E_2(n))$. ELC(n-1)=$(V_2(n-1),E_2(n-1))$, and ELC(n-2)=$(V_2(n-2),E_2(n-2))$, then $V_2(n)= 0V_1(n-1) \cup 10V_1(n-3)0$. For example, V(5) = 0{00,10,11,01} $\cup$ 10{}0 = {000,010,011,001,100}. In other words, ELC(n) can be decomposed into two subgraphs that are isomorphic to ELC(n-1) and ELC(n-3), respectively, for n $\geq$ 5. Figure 5 illustrates ELC(n), n=3,4,5,6.

In this paper, we first show that there exists a Hamiltonian cycle in any ELC. Thus, there is a dilation 1 embedding of a linear array and ring into its corresponding optimum ELC.

Then, we consider the problem of simulating extended Fibonacci meshes on EFC. Finally using these results, we describe a dilation 1 embedding of the extended Fibonacci meshes ( An m × k mesh is a extended Fibonacci mesh iff m = $f_i$ and k = $F_j$ for some i and j) into their optimum Extended Lucas cube.

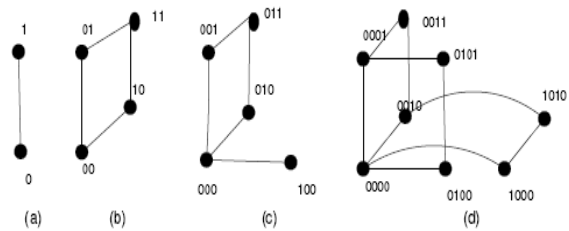

**Fig. 5. a) ELC(3), b) ELC(4), c) ELC(5) , d) ELC(6)**

## 3. Embeddings of Linear Arrays and Rings

**Lemma 1:** For any n $\geq$ 2, n $\neq$ 3, ELC(n) is a Hamiltonian graph.

*Proof :* For n=2 , ELC(n) is a Hamiltonian graph; it has a Hamiltonian cycle : 00→10→11→01. For n=3, ELC(3) is not a Hamiltonian graph. However, it has a Hamiltonian path : 100→000→001→011 →010. Using the induction technique can be proved that ELC(n) is a Hamiltonian graph. For each ELC(n), n $\geq$ 4, a Hamiltonian path of type $000^{n-2} \rightarrow 100^{n-2} \rightarrow G \rightarrow 010^{n-2}$ can be constructed, where $0^{n-2}$ is a sequence of 0 of length n-2 (could be empty). G is a sequence of adjacent to $100^{n-2}$ and $010^{n-2}$ , respectively. For n = 4, a Hamiltonian can be obviously constructed in ELC(4); that is 0000 → 1000 → 1010 → 0010 → 0011 → 0001 → 0101 → 0100. Since the first and last vertices of this Hamiltonian path are adjacent, this Hamiltonian path is also a Hamiltonian cycle. As it can be seen in [7], ELC(n) contains two disjoint subgraphs are isomorphic to EFC(n-1) and EFC(n-3) respectively. In addition, the code for EFC(n) is two bit longer than the one for EFC(n-2). That the theorem holds for n = 2,3,4 is evident . Furthermore, based on those above, with using the induction technique it can be proved that ELC(n) is a Hamiltonian graph. □

**Theorem 2:** A linear array can be embedded into its corresponding optimum ELC with dilation 1.

*Proof :* Immdiately from Lemma 1.

**Theorem 3:** Ring network R(b) can be embedded into ELC(n) with dilation 1, b = $2.f_n \times 2.f_{n-2}$

*Proof :* Immediately from Lemma 1. □

## 4. Embeddings of Extended Fibonacci Meshes

In this section, we first consider the problem of simulating extended Fibonacci meshes on EFCs, then the problem of simulating extended Fibonacci meshes on ELCs

**Theorem 4 :** An $f_n \times F_{n+1}$ extended Fibonacci mesh can be embedded into its corresponding optimum extended Fibonacci cube EFC(2n) with dilation 1.

*Proof* : we know that a linear array with $f_n$ nodes, denoted by LA($f_n$), can be embedded into FC(n) with dilation 1. The 2-D mesh $f_n \times F_{n+1}$ can be considered as matrix with $f_n$ rows and $F_{n+1}$ columns. The basic idea of embedding here is to label columns according to the embedding of LA($F_{n+1}$) to EFC(n+1) and to label rows according to the embedding of LA($f_n$) to FC(n). Thus, the label for the node in the jth column and the ith row in the mesh will be the concatenation of the label for the jth and the ith row. Since a simple concatenation may result in two consecutive 1's, we place one 0 in front of labels for all rows. Clearly, this is a dilation 1 embedding. Furthermore, (n-1) bits are used for labeling columns ( as FC(n+1) has (n+1)-2 bits) and (n-2) +1 bits are used for labeling rows. Thus, total (2n-2) bits are used and this gives us a dilation 1 embedding of the $f_n \times F_{n+1}$ mesh into EFC(2n). Figure 6 illustrates this technique, and Figure 7 shows an example. Now, we show that EFC(2n) is the smallest extended Fibonacci cube with at least $f_n \times F_{n+1}$ nodes, [8,9] that is $f_n \times F_{n+1} > |EFC(2n-1)| = F_{2n-1}$.

$f_n \times F_{n+1} = f_n \times (F_n + F_{n-1}) = f_n F_n + f_n F_{n-1} = f_n F_n + (f_{n-1} + f_{n-2}) F_{n-1} = f_n F_n + f_{n-1} F_{n-1} + f_{n-2} F_{n-1}$

Since $f_n F_n + f_{n-1} F_{n-1} = F_{2n-1}$, we have

$f_n \times F_{n+1} = F_{2n-1} + f_{n-2} F_{n-1} > F_{2n-1} = |EFC(2n-1)|$ □

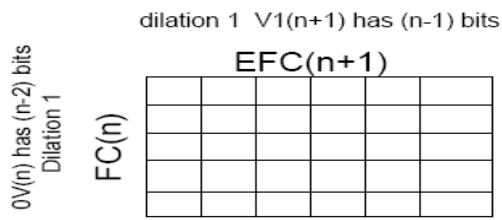

**Fig. 6.** Embedding extended Fibonacci meshes into EFC

For dilation 1 embeddings of square meshes into the EFCs, we have the following result :

**Corallary 5 :** An $f_n \times F_n$ extended Fibonacci mesh can be embedded into its optimum extended Fibonacci cube EFC(2n-1) with dilation 1.

*Proof :* Obtained by using the same technique illustrated in the proof of Theorem 4. □

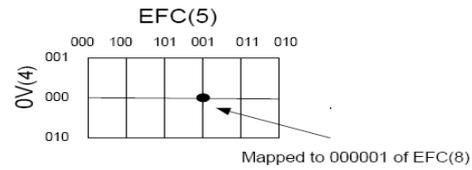

**Fig.7 :** Embedding of $f_4 \times F_5$ mesh into EFC(8).

**Theorem 6 :** Two disjoint subgraphs $f_n \times F_n$ extended Fibonacci mesh and $f_{n+1} \times F_{n+1}$ extended Fibonacci mesh can be embedded directly into EFC(2n+1) with dilation 1. Two disjoint subgraphs $f_n \times F_{n+1}$ extended Fibonacci mesh and $f_n \times F_{n-1}$ extended Fibonacci mesh can be embedded into EFC(2n) with dilation 1.

**Theorem 7 :** Two disjoint subgraphs $f_n \times F_{n+1}$ extended Fibonacci mesh and $f_n \times F_{n-1}$ extended Fibonacci mesh can be embedded into its corresponding optimum extended Lucas cube ELC(2n+1) with dilation 1.

*Proof :* We know that two disjoint subgraphs EFC(2n) and EFC(2n-2) can be embedded directly into ELC(2n+1). From Theorem 6, two disjoint subgraphs $f_n \times F_{n+1}$ extended Fibonacci mesh and $f_n \times F_{n-1}$ extended Fibonacci mesh can be embedded into EFC(2n). Since $|EFC(2n)| > |EFC(2n-2)|$, then the theorem are proved. □